\documentclass[prl,superscriptaddress, notitlepage]{revtex4-1}

\usepackage[utf8]{inputenc} 
\usepackage[T1]{fontenc}    
\usepackage{hyperref}       
\usepackage{url}            
\usepackage{booktabs}       
\usepackage{amsfonts}       
\usepackage{nicefrac}       
\usepackage{microtype}      
\usepackage{lipsum}		
\usepackage{graphicx}
\usepackage{natbib}
\usepackage{lmodern} 
\usepackage{microtype}
\usepackage{natbib}

\begin{document}

\title{Lithography-free directional control of thermal emission.}
\author{Mitradeep Sarkar}

\author{Maxime Giteau}
\author{Michael Enders}
\author{Georgia T. Papadakis}
\email{mitradeep.sarkar@icfo.eu}
\affiliation{
ICFO-Institut de Ciencies Fotoniques, The Barcelona Institute of Science and Technology,\\ 08860 Castelldefels (Barcelona), Spain
}

\begin{abstract}
	Blackbody thermal emission is spatially diffuse. Achieving highly directional thermal emission typically requires nanostructuring the surface of the thermally emissive medium. The most common configuration is a subwavelength grating that scatters surface polaritonic modes from the near-field to the far-field and produces antenna-like lobes of thermal emission. This concept, however, is typically limited to a particular linear polarization, and requires sophisticated lithography. Here, we revisit the simple motif of a planar Salisbury screen. We show analytically how the interplay between the real and imaginary part of the dielectric permittivity of the emitter layer defines the directional characteristics of emission, which can range from diffuse to highly directional. We propose a realistic configuration and show that hexagonal Boron Nitride thin films can enable grating-like thermal emission lobes in a lithography-free platform.\\
\textbf{Nanophotonics:}  Thermal Emission, Mid-Infrared , Phonon Polaritons
\end{abstract}

\maketitle

\section{Introduction} 
Mid-infrared (IR) thermal emission is ubiquitous in nature. Gaining control over its spatial and spectral characteristics is central to various applications. Examples include light and energy harvesting \cite{Li2018reviewfan, Fan2022}, as in thermophotovoltaic systems \cite{LaPotin2022,Datas2019,Zhao2013,Lenert2014,Sakakibara2019}, radiative cooling \cite{Linxiao2015}, IR sources \cite{Junghyun2022,Hojo2021}, thermal camouflage \cite{Li2018} and molecular sensing \cite{Oh2021biosensors,Barho2019}. Spectrally narrow thermal emission is required in various applications, such as in sensing \cite{Liu2011,Gillibert2016,Lochbaum2017,Sun2021}, IR sources \cite{Campione2016,Wang2018}, as well as for maximizing light-harvesting conversion efficiency \cite{Papadakis2021,Papadakis2020}. In these applications, control over the \textit{spatial} characteristics of thermal emission is key.

\par{Since blackbody thermal emission is spectrally broad and spatially diffuse, photonic design can be employed to narrow its spectral range \cite{DeZoysa2012,Xiang2012} and to control its directionality \cite{Baranov2019}, or both \cite{Sun2021,Greffet2002}. A plethora of sophisticated nanostructures have been considered for tailoring thermal emission, such as mid-IR antennas \cite{Adato2015,Semple2021}, multi-layered films and one-dimensional photonic crystals \cite{Vasanelli2021,Campione2016,Qudirectional}, complex three-dimensional resonators \cite{constantinidirectionalplasmon,Lu2021phononcoupling, MaHuangZhaQinQinGhoshKaurQiuLi+2022}, and gratings \cite{Junghyun2022,Ito2014}. In fact, the principle of operation of many of these motifs and other sophisticated designs \cite{Inoue2014,Wojszvzyk2021,Jin2021, He2022} lies within the physics of a photonic grating. As Greffet \textit{et al.} experimentally demonstrated in \cite{Greffet2002}, one can thermally excite near-field surface phonon polariton modes \cite{Caldwell2015} on a silicon carbide (SiC) surface. Using a subwavelength grating, it was shown that these modes can diffract into propagating modes at specific angles, thus enabling an ultra-high degree of directional control in thermal emission. Such surface phonon polaritons occur strictly within the Reststrahlen band of SiC, which is the frequency range where its dielectric function is negative \cite{gallgreffet1997}. The effect can be generalized to any material with a polar resonance in the mid-IR  \cite{surfacepolaritonIR}, as well as to plasmonic media for frequencies below their plasma frequency, where they support surface plasmon polaritons \cite{Scholl2012,Sarkar2015}}.

In both plasmonic and polar media, however, surface polaritons occur solely for p-polarization \cite{Papadakis2018,georgiaplasmon_OP}, thus the aforementioned concept is also constrained to p-polarized radiation. Furthermore, the Reststrahlen band where surface phonon polaritons occur is spectrally narrow \cite{Caldwell2015}, thus limiting the frequency range of operation of a grating and related motifs for thermal emission control. Importantly, nanoscale patterning requires complex nanolithography or synthesis.

\begin{figure*}[]
	\centering
	\includegraphics[width=1\linewidth]{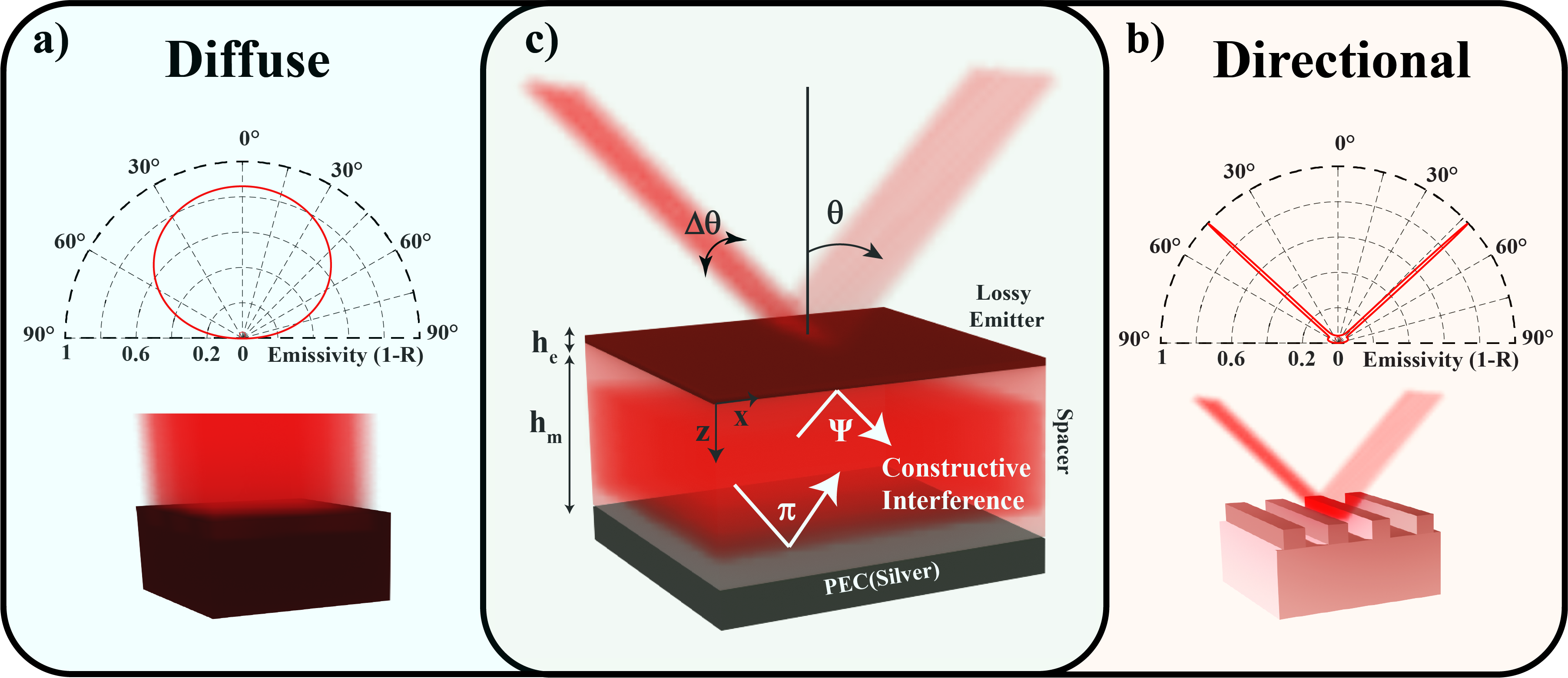}
	\caption{a) Black-body radiation can be classified as diffuse and the polar plot of a typical black-body emissivity is shown. b) Periodic nano-structures (gratings) have directional emission at well defined narrow spectral ranges. c) The schematic of the meta-structure comprising of a lossy emitter (permittivity $\epsilon_e$) and a lossless spacer (permittivity $\epsilon_m$) in the Salisbury screen configuration. Light in the spacer will acquire a phase of $\pi$ for reflection from the PEC and $\Psi$ on reflection from the lossy emitter. When the phase matching condition (constructive interference in the structure) is satisfied, the structure has high emissivity. Directional thermal emission at a particular angle $\theta$ with angular spread $\Delta\theta$ is achieved when specific interference conditions are satisfied.}
	\label{fig:1schematic} 
\end{figure*}

In contrast to previous attempts to control the directionality of a thermally emitted beam via a nanostructure, here, we revisit the concept of a Salisbury screen \cite{salisbury_1952}. A Salisbury screen is a three-layer planar structure that operates based on constructive interference between the layers, and yields near-unity thermal emissivity on resonance \cite{Fang2018dalenbachsalisbury}. It is constructed out of a thin thermal emitter on a thick dielectric spacer with a back-side reflector, as shown in Fig.\ref{fig:1schematic}c. We study the directionality of thermal emission from a Salisbury screen in a systematic and analytical way, with respect to the optical properties of the emitting layer. By considering two regimes of thermal emission: spatially diffuse and directional, we present simple nanophotonic strategies to achieve both, in a lithography-free configuration. We show that a key characteristic for highly directional thermal emission is a high-absolute value of the dielectric permittivity ($\epsilon$) of the emitter layer, which can be found in various polar media \cite{Caldwell2015}. As a practical example, we leverage the high-dielectric permittivity of hexagonal Boron Nitride (hBN) near 7$\mu m$ to demonstrate that grating-like thermal emission lobes that are highly directional, without \textit{any} lithography, for \textit{both} linear polarizations.

We note that recent works have reported directional thermal emission in similarly planar configurations \cite{baranov_semi_inf,Jin2021,Lee2006,Johns2022}. However, in \cite{Lee2006}, the photonic band gap of photonic crystals was used to achieve strong directionality, which requires a very large number of layers. In addition, it was shown that Salisbury-like heterostructures can yield directionality, however the concept was limited to metal-spacer-metal structures, hence much weaker performance. In \cite{baranov_semi_inf}, by employing strong optical anisotropies, it was shown that one can use the Brewster's condition to achieve directionality, however this concept pertains to near-grazing angles of incidence. In \cite{Jin2021}, broadband directional emission was experimentally demonstrated, whereas in \cite{Johns2022}, it was theoretically shown that epsilon-near-zero films on a reflector can direct a thermally emitted beam for p-polarized light. Nevertheless, in all the above, the reported directionality remains significantly inferior to that of a grating \cite{Greffet2002}.

\section{Directional control with planar meta-structure} 

The directionality of a thermally emitted beam can be described by the dependence of the thermal emissivity, $\mathcal{E}(\theta)$, on the zenith angle of emission, $\theta$. From Kirchhoff's law of thermal radiation \cite{Kirchhoff1860}, the emissivity is equal to the absorptivity, $\mathcal{E}=\mathcal{A}$. For opaque emitters, therefore, $\mathcal{E}=1-R$, where $R$ is the reflectivity of the emitter. In order to evaluate the performance of various geometries as directional thermal emitters at a particular wavelength and linear polarization (namely s and p, pertaining to transverse electric and transverse magnetic waves, respectively), with respect to a polar diagram such as those in Figs. \ref{fig:1schematic}(a)-(c), we consider two relevant properties: (i) the angular spread of the emissivity, $\Delta\theta$, which is defined here as the full width half maximum (FWHM) of the lobe, and (ii) the contrast between maximum and minimum emissivity at this wavelength, defined as $C=\left|R^{min}(\theta)-R^{max}(\theta)\right|$. The ratio $C/\Delta\theta$ expresses the degree of directionality of a thermal beam. Since we can write $\Delta\theta\approx|dR/d\theta|^{-1}$, we define a figure of merit of the directionality of a thermal beam ($\mathrm{FOM}$) as
\begin{equation}\label{eq:FOM_define}
	\mathrm{FOM}=C\times\left|\frac{dR}{d\theta}\right|.
\end{equation}
By the definition of Eq. \ref{eq:FOM_define}, a low $\mathrm{FOM}$ indicates diffuse (unidirectional) thermal emission, such as a the spatial characteristic of a thermal emission from a blackbody, whereas a large $\mathrm{FOM}$ suggests highly directional emission, such as that achieved by a the grating \cite{Greffet2002}. These are shown in Figs. \ref{fig:1schematic} (a), (b), respectively.
In the following section, we will discuss a strategy to achieve both narrowband diffuse and narroband directional thermal emission with a lithography-free planar configuration, the Salisbury screen \cite{salisbury_1952}. In this configuration, a dielectric spacer layer is sandwiched between a thin lossy emitter and a perfect electric conductor (PEC) as shown schematically in Fig.\ref{fig:1schematic} (c). The thickness of the dielectric spacer is a multiple of approximately $\lambda_\mathrm{o}/4$, where $\lambda_\mathrm{o}$ is the free-space wavelength, to warrant constructive interference at the air-emitter interface, for maximal thermal emission.

\subsection{The phase matching condition.}

To have minimum reflectivity from the structure at a given incidence angle $\theta$, it is necessary to have constructive interference of light in the spacer. Considering an acquired phase of $\pi$ and $\Psi$ for reflections at the PEC and lossy material respectively, the condition for constructive interference in the structure (phase-matching condition) can be written as \cite{Lee2006}
	
	\begin{equation}\label{eq:phase matching}
		\begin{array}{c c l}\displaystyle
		2k_0h_m\aleph_m+\pi+\Psi=2l\pi	\\\\
		\textbf{Directional} : \Psi\approx\pi\\\\
	\textbf{Diffuse} : \Psi\approx 0,2\pi\\
			\end{array}
			\end{equation}
	
	\par{ where $l$ is an integer denoting the interference order and $k_0=2\pi/\lambda$, with $\lambda$ the wavelength of light.}
	\par{The emitter (with subscript $e$) and the spacer (with subscript $m$) can be considered anisotropic, having a diagonal permittivity tensor with elements $\epsilon$, $\epsilon/\alpha_{Y}^2$ and $\epsilon/\alpha_{Z}^2$ along $x$, $y$ and $z$ axes respectively. We introduce the terms $\aleph_Z=\sqrt{\epsilon-\alpha_Z^2sin^2\theta}$ and $\aleph_Y=\sqrt{\epsilon-\alpha_Y^2sin^2\theta}/\alpha_Y$ and call them the effective index of the anisotropic medium along $z$ and $y$ axes respectively. These terms are used throughout the article to simplify the analytical expressions for each media with the respective medium subscript.}
	
	\par{Considering the refractive index of the spacer to be close to unity and for the first interference order ($l=1$), we can rewrite Eq.\ref{eq:phase matching} as $\cos \theta=(\pi-\Psi)/(2k_0h_m)$. Hence it is clear that if we need high directional control (stronger $\theta$ dependence of the phase matching condition), we need the RHS close to zero so that the cosine function has the maximum gradient with respect to $\theta$. So, for maximum directional control, $\Psi=\pi$. The inverse is true as well, hence for minimum angular dependence, (diffuse thermal emission), we need $\Psi=0,2\pi.$. These two conditions as discussed intuitively above is shown in Eq.\ref{eq:phase matching}. So a knowledge of $\Psi$ will allow us to define the conditions required for directional control.}
	
	\par{We should note that the angular dependence of the phase matching condition stems from the term $\aleph_m=\sqrt{\epsilon_m-sin^2\theta}$ (considering weak $\theta$ dependence of $\Psi$), and this $\theta$ dependence decreases as $\epsilon_m$ becomes larger than unity. Hence for stronger directional emission, the spacer should have a low refractive index.}
	
	\par{To derive $\Psi$, the reflectivity ($R_{P/S}$) from the meta-structure was calculated for both transverse magnetic (P) and transverse electric (S) polarizations. We also know that when the reflectivity is zero, the phase matching condition of Eq.\ref{eq:phase matching} is satisfied. Hence, the acquired phase on reflection from the lossy material ($\Psi$) can be derived by comparing Eq.\ref{eq:phase matching} and the condition for $R_{P/S}=0$. This can be shown to be }
	\begin{equation}\label{eq:phase PSI}
		\Psi_P=\pi-2tan^{-1}\frac{\aleph_{Ze}\epsilon_m}{\epsilon_e\aleph_m}\frac{\epsilon_ecos\theta+i\aleph_{Ze}T_e}{\epsilon_ecos\theta T_e-i\aleph_{Ze}} 
	\end{equation}
	\par{for P polarization. A similar equation can be derived for the S polarization.}
	
	\par{We see from Eq.~\ref{eq:phase PSI} that when $h_e=0$, $\Psi=0$ and the phase matching condition simply becomes $h_m=(2l-1)\lambda/4\aleph_{m}$ which is the condition for constructive interference in the spacer of a Salisbury screen configuration when the phase from the emitter can be neglected. Thus, without the lossy-emitter, the thermal emission is always expected to be diffuse.}

\begin{figure*}[]
	\centering
	\includegraphics[width=0.8\linewidth]{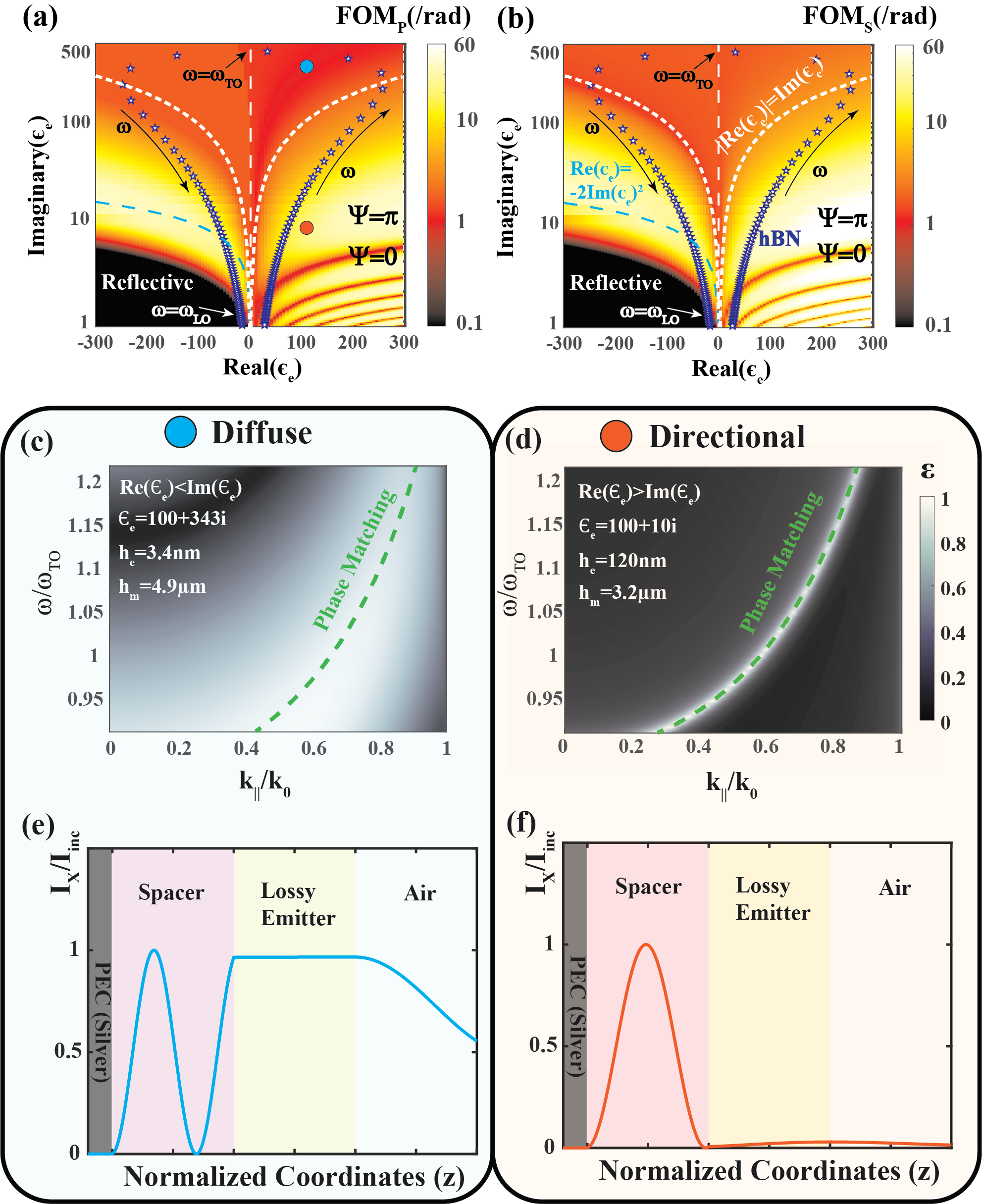}
	\caption{FOM of the meta-structure as a function of the real and imaginary parts of $\epsilon_e$ for P (a) and S (b) polarizations. The emitter thickness $h_e$ is fixed according to the critical coupling condition and the spacer thickness is fixed by phase matching condition for each value of $\epsilon_e$ such that the angle of emission $\theta_{em}=45^{\circ}$. The values of the permittivity of hBN are superposed on the figure for frequencies ($\omega$) close to the Reststrahlen band. We can separate the map into two regimes: High FOM (directional) when $\Psi\approx\pi$ and low FOM (diffuse) when $\Psi\approx 0,2\Pi$ (c,d) The emissivity ($\mathcal{E}_P=1-R_P$) as a function of $\omega$ (normalized to $\omega_{TO}=1360cm^{-1}$ of hBN) and in-plane wave-vector $k_{||}/k_0=sin\theta$ for fixed $\epsilon_e$, $h_e$ and $h_m$ for diffuse ($Re(\epsilon_e)<Im(\epsilon_e)$) and directional ($Re(\epsilon_e)>Im(\epsilon_e)$) emission. The values of $\epsilon_e$ for each is shown in (a) as color dots. The phase matching condition calculated from Eq.\ref{eq:phase matching} and Eq.\ref{eq:phase PSI} is superposed as green dashed lines. (e,f)  The corresponding in-plane electric field intensity for the same $\epsilon_e$ as in c and d respectively. The coordinates of each layer (emitter, spacer) are normalized by their thicknesses ($h_e$, $h_m$) respectively.}
	\label{fig:2FOMana} 
\end{figure*}
	
\par{Now let us define the geometrical parameters for the meta-structure. For the emitter thickness $h_e$, we will use the critical coupling condition. From coupled-mode theory the emissivity is unity at a resonant frequency, when the radiative decay rate ($d\mathcal{E}/dt$) equals the non-radiative decay rate which in turn is equal to half the damping rate of the material \cite{raman_radiativeloss}. So to have $\mathcal{E}=1$, we need $h_e=\lambda /(2\pi \Im(\epsilon_e))$ as shown in Zhao et al~\cite{zhao_hbn}.}

\par{Using the above expression for emitter thickness and the phase matching condition, the spacer thickness $h_m$ can be calculated for any given angle of emission $\theta_{em}$. }

\par{It is interesting to mention here that with the above geometrical parameters for $\epsilon_m=1$ and $\theta=0$, we can approximate the phase acquired on reflection from the lossy material to be $\Psi_P=\pi-2\arg({\epsilon_e})$. The same can be shown for TE polarization. We can simply see from this relation that for $Re(\epsilon_e)\gg Im(\epsilon_e)$, $arg(\epsilon_e)=0$, so $\Psi=\pi$ while for $Re(\epsilon_e)\ll Im(\epsilon_e)$, $\arg(\epsilon_e)\approx \pi/2$ and $\Psi\approx 0$. This condition relating $\Psi$ to the real and imaginary parts of $\epsilon_e$ is crucial and is demonstrated next.}	

\subsection{The Figure of Merit for meta-structures}

\par{The $FOM$ for both polarizations was calculated numerically using the expressions for $R_{P/S}$ and Eq.\ref{eq:FOM_define} and is shown Fig.\ref{fig:2FOMana} as a function of the real and imaginary parts of $\epsilon_e$. The emitter thickness was fixed by the critical coupling condition while the spacer thickness was obtained by the phase matching condition for each value of complex $\epsilon_e$ to have maximum emission at $\theta_{em}=45^{\circ}$. The spacer permittivity was fixed at $\epsilon_m=1.65$.}

\par{We clearly see two distinct zones where the $FOM$ is high, on two sides of $Re(\epsilon_e)=0$ (Epsilon near zero, ENZ). So for high FOM, we need high negative or positive values of the real part of the permittivity. This condition can be satisfied over a specific frequency range close to the Reststrahlen band in hBN (permittivity shown as stars in Fig.\ref{fig:2FOMana}a,b). }	
	
\begin{figure*}[]
	\centering
	\includegraphics[width=1\linewidth]{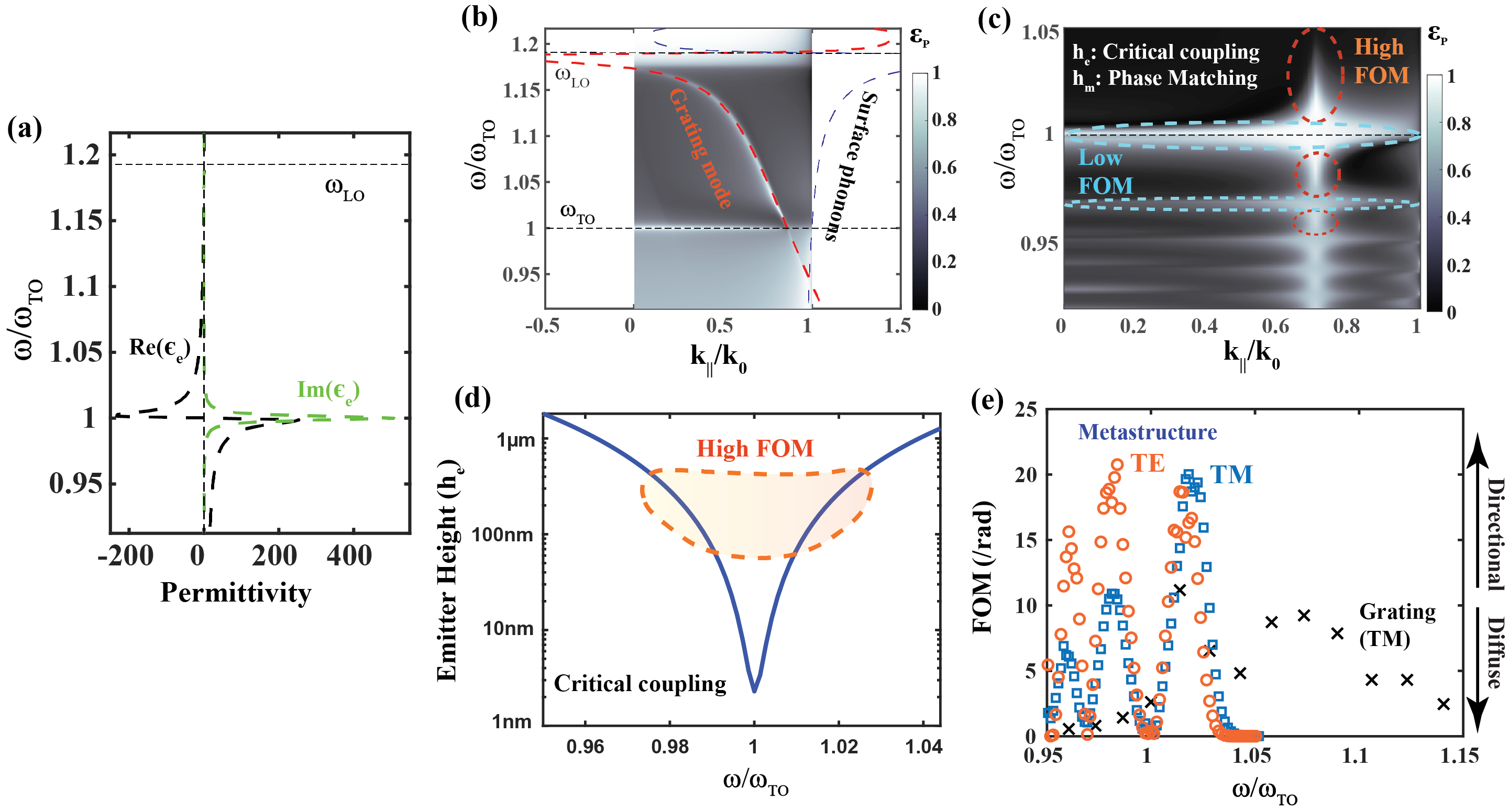}
	\caption{(a) The in plane permittivity of hBN around the in-plane Reststrahlen band ($\omega_{TO,\perp}=1360cm^{-1}$). (b) $\mathcal{E}_P$ for a semi-infinite hBN grating with grating height of $250nm$ and period $3.92\mu m$. The dispersion of the surface phonons $k_{SP}$ (black dashed) and the grating condition (red dashed) is shown. (c) $\mathcal{E}_P$ for the meta-structure with hBN emitter, LiF spacer and silver substrate. The emitter thickness $h_e$ is selected by the critical coupling condition for each $\omega$ and shown in (d). The spacer thickness $h_m$ is selected by the phase matching condition with $\theta_{em}=45^{\circ}$. (e) The FOM for P polarization (squares) and S polarization (circles) for the meta-structure shown in c. The FOM for the grating of (b) is also shown (black crosses).}
	\label{fig:3hBN} 
\end{figure*}

\par{Let us now discuss the different aspects of the FOM map. It can be shown analytically that the FOM will be high when $Re(\epsilon_e)>Im(\epsilon_e)$ and for $Re(\epsilon_e)=-2Im(\epsilon_e)^2$. These correspond to the two zones of high FOM as shown in Fig.\ref{fig:2FOMana}a,b.}

\par{The dispersion of the emissivity ($1-R_{P}$) for fixed values of $\epsilon_e$, $h_m$ and $h_e$ are shown in Fig.~\ref{fig:2FOMana}(c,d). We see that the phase matching condition as given by Eq.~\ref{eq:phase matching} with $\Psi$ calculated by Eq.~\ref{eq:phase PSI} corresponds perfectly the positions of maximum emissivity. We can note that we have two regimes of angular control, depending on the values of complex $\epsilon_e$. If $Re(\epsilon_e)>Im(\epsilon_e)$ we have directional emission while it is diffuse when $Re(\epsilon_e)<Im(\epsilon_e)$. For the geometrical parameters used in these figures and using Eq.\ref{eq:phase PSI}, we have seen that $\Psi=\pi$ for the directional regime and $\Psi=0$ for the diffuse regime, as expected by the intuitive analysis presented before.}

\par{Another interesting point to note is that for high values of $\Im(\epsilon_e$), even though the emissivity is high, the FOM, hence the directional control achieved, is extremely low. We also see the FOM is close to zero for certain values of complex $\epsilon_e$ (in the positive $\Re(\epsilon_e)$ part of the FOM map). This is because for strong directional control of emissivity, the two interfaces of the spacer must be symmetric in terms of acquired phase ($\Psi=\pi$). However, for certain values of $\epsilon_e$, there can be constructive interference in the lossy emitter and in these cases $\Psi=0$ and we have diffuse emission and low $FOM$. Also for diffuse emission, we need $Re(\epsilon_e)<Im(\epsilon_e)$ which cannot be satisfied by metals in the mid IR, so emission from metal-spacer-metal structures are always directional. }

\par{The above discussion is justified further from the field profiles shown in Fig.~\ref{fig:2FOMana}(e,f). Regions of high FOM, hence lower angular spread correspond perfectly with the situation when the field is maximum at the center of the spacer, while almost zero near the spacer-emitter interface (symmetric to the spacer-PEC interface). However, if the field is high in the emitter (constructive interference in the emitter), the angular control of thermal emission is drastically reduced as the field distribution is no longer symmetric on the two sides of the spacer.}
\par{In this section our discussions were based on the P polarized light. The same can be shown for the S polarized light. Now we will present these design ideas with respect to hBN as the emitter. Our results can be generalized to any material which can support polaritons and hence have a complex permittivity.}

\section{Meta-structure with hexagonal Boron Nitride} 

\begin{figure}[]
	\centering
	\includegraphics[width=0.5\linewidth]{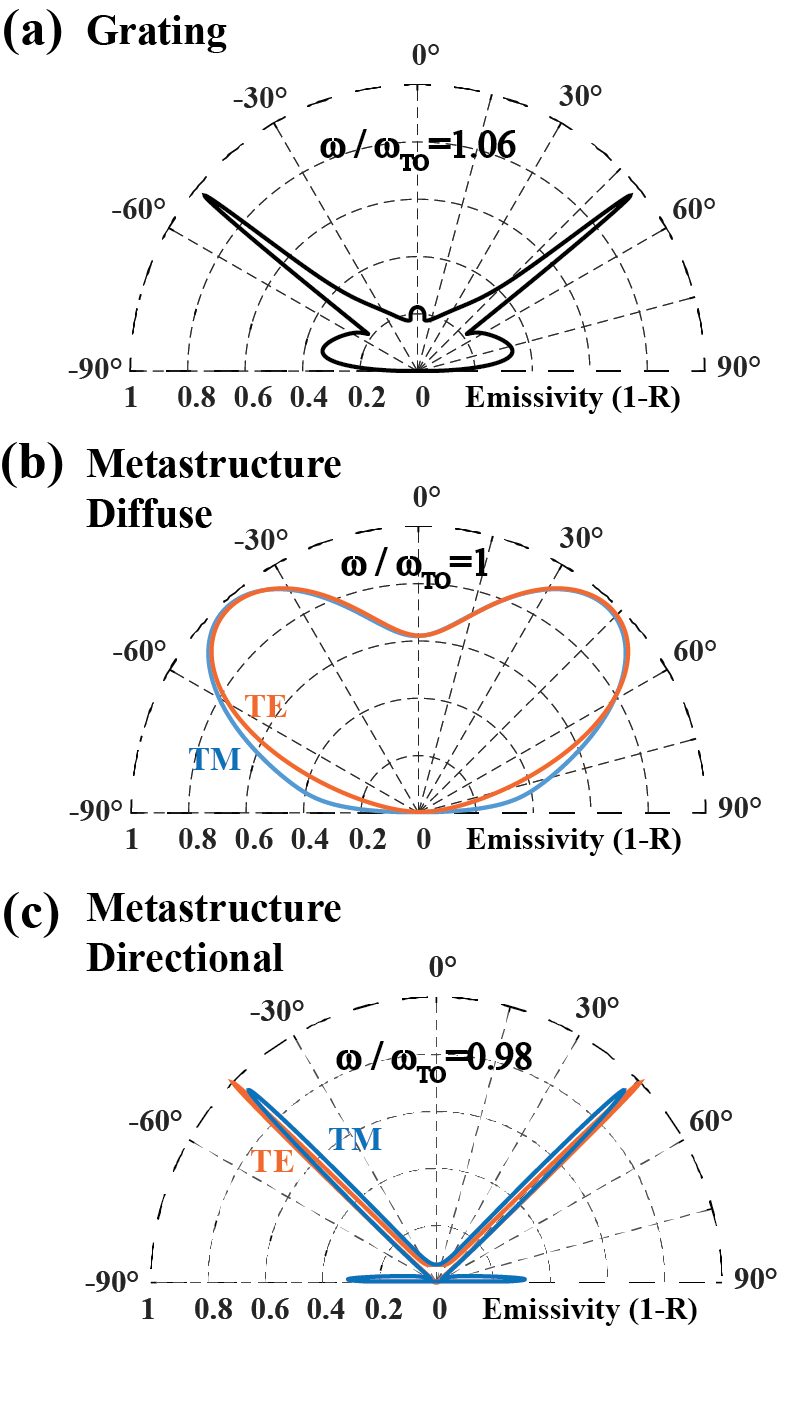}
	\caption{(a) The polar plot for P polarized emission from the grating structure shown in Fig.\ref{fig:3hBN}b for a frequency in the center of the Reststrahlen band of hBN. (b) The polar plot of $\mathcal{E}$ for the meta-structure when $\omega=\omega_{TO}$ (diffuse emission) for both P and S polarizations. (c) The same for a frequency where $Re(\epsilon_e)>Im(\epsilon_e)$, hence directional emission for both polarizations.}
	\label{fig:4polarplots} 
\end{figure}	

\par{Hexagonal Boron Nitride (hBN) has two Reststrahlen bands: an in-plane phonon mode at around 7 $\mu$m and an out-of-plane phonon mode at 13 $\mu$m \cite{Caldwell2014hbn}. We will focus on the in-plane modes as the permittivity along $x$ or $y$ (in-plane) plays a major role for our purpose. We use the bulk model of hBN for which the permittivity can be modeled as }

\begin{equation}
	\label{eq:hBNperm}
	\epsilon_e=\epsilon_{inf}\left[1+\frac{\omega_{LO}^2-\omega_{TO}^2}{\omega_{TO}^2-i\Gamma\omega-\omega^2}\right]
\end{equation}

\par{with in plane parameters ($x,y$) being $\omega_{L0,\perp}=1610 cm^{-1}$, $\omega_{T0,\perp}=1360 cm^{-1}$, $\Gamma_{\perp}=5cm^{-1}$ and $\epsilon_{inf,\perp}=4.87$. The out of plane parameters ($z$) are $\omega_{L0,||}=830 cm^{-1}$, $\omega_{T0,||}=780 cm^{-1}$, $\Gamma_{||}=4cm^{-1}$ and $\epsilon_{inf,||}=2.95$ \cite{zhao_hbn}. The in-plane permittivity of hBN is shown in Fig\ref{fig:3hBN}(a) for frequencies around the in-plane Reststrahlen band.}
\par{In actual experiments, that deviations from the above parameters will not effect the basic finding of this article. The geometrical parameters will then have to be tuned according to the critical coupling/phase matching conditions depending on the actual values of $\epsilon_e$. High FOM is achieved close to $\omega_{TO}$ for frequencies for which $|Re(\epsilon_e)|>Im(\epsilon_e)$ is satisfied and the design must be done accordingly. Also higher absolute values of $Re(\epsilon_e)$ can be achieved if $\Gamma$ decreases hence that will result in high FOM.}

\par{The spacer layer is considered to be Lithium Fluoride (LiF) which has low dispersion over the mid IR with a refractive index of 1.28 for $\lambda = 7.3$ µm. As mentioned before, for maximum contrast and minimum angular spread, $\epsilon_m$ should be close to unity (permittivity of the incidence medium) as possible to ascertain better confinement of the light in the spacer for lower orders $l$. The spacer needs to be a few micrometers thick for effective spatial control, hence it is imperative that the spacer is lossless ($h_m\ll$skin depth of material).}
\par{The numerical calculations for the reflectivity of all structures (gratings and planar) is done by the Rigorous coupled wave analysis (RCWA) package by Hugonin and Lalanne \cite{hugonin2021reticolo}.}

\par{The dispersion of the surface phonon for grating coupling is shown in Fig.~\ref{fig:3hBN}(b). The surface phonon mode can be excited when the grating condition $k_0 \sin\theta+k_{SP}-2\pi/P=0$ is satisfied \cite{SarkarOE2015} for grating period $P=3.92 \ \mu m$ and the in-plane wave-vector of the surface phonons being $k_{SP}=k_0\sqrt{[\epsilon_e-\epsilon_e^2]/[\alpha_{Ze}-\epsilon_e^2]}$. The grating height is taken to be 250 nm which is close to the penetration depth of the surface phonons in air. The maximum emissivity is achieved for all frequencies lying between $\omega_{LO}$ and $\omega_{TO}$. It should be noted that the emissivity is not unity for the grating and this is due to hBN being anisotropic. }

\par{The complete spectral and spatial map for the emissivity of the meta-structure is shown in Fig.~\ref{fig:3hBN}(c). For this result, the maximum emission angle ($\theta_{em}$) was chosen to be $45^{\circ}$ with respect to the TM polarization and the spacer thickness was varied accordingly for each frequency via the phase matching condition for $l=1$ (for $\omega>\omega_{TO}$) and $l=2$ (for $\omega<\omega_{TO}$). The emitter thickness was fixed for each frequency according to the critical coupling condition and is shown in Fig.\ref{fig:3hBN}(d).}
\par{The narrow emissivity peaks, similar to the grating at $\theta=45^{\circ}$ ($k_{||}/k_0=0.707$) in Fig\ref{fig:3hBN}c demonstrates that strong spatial control can be achieved for lithography-free planar meta-structures when operating close to the Reststrahlen band edge ($\omega_{TO}$) of polar materials (for conditions when $\Psi\approx\pi$). The horizontal bands in the emissivity maps at certain frequencies where the thermal emission is diffuse are same as the low FOM bands in Fig.\ref{fig:2FOMana}a,b where the constructive interference occurs in the emitter instead of the spacer ($\Psi=0$).  }

\par{The FOM for the meta-structure was calculated numerically and is shown in Fig.~\ref{fig:3hBN}(e). We see the two distinct zones of high FOM on either side of $\omega_{TO}$ as discussed in the previous section. Also for $\omega=\omega_{TO}$, when $Re(\epsilon_e)\ll Im(\epsilon_e)$, we have diffuse thermal emission ($FOM=0$). Furthermore, the FOM does not depend strongly on $\theta_{em}$ and the meta-structure can be optimized for any emission angle by changing the spacer thickness $h_m$.} 

\par{The FOM for the meta-structure is compared with the FOM of the anisotropic hBN grating. The FOM for an isotropic grating will be roughly two times larger than what is shown, hence similar to the FOM achieved with the planar meta-structure.}
\par{The polar plots of emissivity is shown for the anisotropic grating, and the two functionalities of the meta-structure for both polarizations in Fig\ref{fig:4polarplots}. Thus we can achieve directional emissivity lobes as narrow and with high contrast as obtained by the grating, however for both polarizations and without the requirement of lithography. We can also switch between diffuse and directional emission, depending on the frequency of emission and the geometrical parameters chosen for the structure.}

\section{Conclusions} 
\par{In this article we presented a method to bypass the inherent difficulty in achieving strong spatial and spectral control of thermal emission. Furthermore, unlike most other methods, our configuration can achieve the control for both polarizations simultaneously. We present a simple geometry involving a Salisbury screen configuration and by intelligent tailoring of the emitter and spacer thicknesses, we can strongly confine light emitted due to the volume polaritons of polar materials in the lossless spacer. This results in strong directional control of the emitted light, which can be passively tuned. Our structure can satisfy the conditions required for both directional and diffuse thermal emission. Our analytical calculations were validated by rigorous numerical calculations for actual materials.The structure presented here does not require expensive lithography or fabrication techniques and can be easily adapted for various applications which require such strong spatial and spectral control.}

\section*{Acknowledgments}
The authors declare no competing financial interest. The project that gave rise to these results received the support of a fellowship from ”la Caixa” Foundation (ID 100010434) and from the European Union’s Horizon 2020 research and innovation programme under the Marie Skłodowska-Curie grant agreement No 847648. The fellowship code is LCF/BQ/PI21/11830019. M.Giteau acknowledges financial support from the Severo Ochoa Excellence Fellowship.

\bibliographystyle{plain}

\end{document}